\documentclass[letterpaper]{article} 
\usepackage{aaai2027}  
\usepackage[hyphens]{url}  
\usepackage{graphicx} 
\usepackage{natbib}  
\usepackage{caption} 
\usepackage{algorithm}
\usepackage{algorithmic}

\usepackage{newfloat}
\usepackage{listings}
\DeclareCaptionStyle{ruled}{labelfont=normalfont,labelsep=colon,strut=off} 
\floatstyle{ruled}
\newfloat{listing}{tb}{lst}{}
\floatname{listing}{Listing}

\usepackage{booktabs}
\usepackage{amsmath}

\title{SkillRefine: Cross-Source Skill Induction and Execution Validation\\
for LLM Agents in Refinery Planning Software}

\author{
    Dongxiao Liu\textsuperscript{\rm 1,\rm 2},
    Boren Zhu\textsuperscript{\rm 2},
    Yuwen Ding\textsuperscript{\rm 2},
    Linghui Li\textsuperscript{\rm 1},
    Li Lei\textsuperscript{\rm 2},
    Xiaoyong Li\textsuperscript{\rm 1}
}
\affiliations{
    \textsuperscript{\rm 1}Beijing University of Posts and Telecommunications, Beijing, China\\
    \textsuperscript{\rm 2}Sinopec Engineering Incorporation, Beijing, China\\
    liudongxiao@bupt.edu.cn, zhuboren.sei@sinopec.com, dingyuwen.sei@sinopec.com,\\
    lilinghui@bupt.edu.cn, lilei.sei@sinopec.com, lixiaoyong@bupt.edu.cn
}

\begin{document}
\maketitle

\begin{abstract}
Operating industrial planning software such as AspenTech
PIMS (Process Industry Modeling System) requires an LLM
agent to combine structural knowledge, procedural knowledge
from expert records, and constraints revealed only during
execution. These evidence sources are heterogeneous and
individually incomplete: documentation describes tables and
interfaces but omits task-level coordination, expert CASE
records expose multi-table modification patterns without
explicit schema grounding, and execution feedback reveals
latent constraints only when a plan is executed. We present
\textsc{SkillRefine}, a framework that exploits the
Documentation--Practice Gap between documented structure
and expert practice to extract candidate coordination patterns,
ground them against table definitions and COM specifications,
and compile them into progressively disclosed skill packages
with provenance. The library is then refined through label-free
compliance screening, oracle-based match decomposition over
table, row, column, and value dimensions, and signal-conditioned
trajectory attribution for localized repair. We evaluate
\textsc{SkillRefine} on PIMS-Bench, a benchmark built from two
AspenTech PIMS demonstration models, using disjoint construction
and held-out test tasks. On the held-out test set,
\textsc{SkillRefine} achieves absolute component match F1 gains
of 14\%--30\% across four LLM backbones, with the largest gains
on complex multi-table coordination tasks.
\end{abstract}

\section{Introduction}

AspenTech PIMS (Process Industry Modeling System)~\citep{aspentech_pims}
represents refinery planning models through 85 interconnected spreadsheet
tables that are modified using a structured CASE interface. Operating PIMS
requires translating business-level instructions---such as ``shut down the
FCC unit'' or ``raise diesel price to \$148/bbl''---into precise CASE entries
with valid table names, row identifiers, column headers, values, and
terminators. Even a routine request such as changing diesel price requires
knowing that diesel is represented by the row tag DSL and that its price is
modified through the PRICE column rather than an intuitive alternative such
as COST. In operational terms, each instruction must be translated into a set
of structured modifications over table, row, column, and value fields, while
preserving cross-table feasibility.

Consider shutting down the FCC unit for maintenance. In the evaluated Gulf
Coast demonstration model, setting FCC capacity to zero in CAPS alone may
produce an infeasible plan because downstream tables encode related yield and
blending assumptions. A valid coordinated modification may additionally
update SLPR yield rows, BLNMIX eligibility columns, and relevant purchase
limits. Correct execution also requires setting both MIN and MAX in CAPS to
zero, targeting exact row tags rather than natural-language descriptions, and
using column names that match the CASE schema. Although the required
modifications are specific to the model and planning scenario, the example
illustrates a general operational challenge within PIMS: a short business
instruction may correspond to several coupled and schema-sensitive
modifications.

\begin{figure}[t]
\centering
\includegraphics[width=\linewidth]{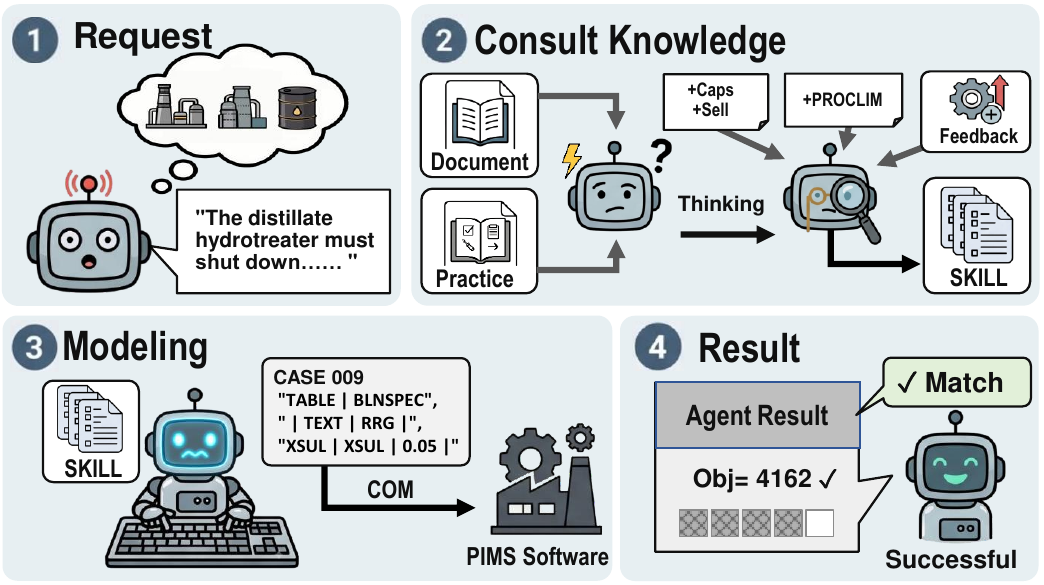}
\caption{Learning operational skills from cross-source knowledge to autonomously operate PIMS.}
\label{fig:problem}
\end{figure}

The knowledge required for these operations is distributed across three
sources, each of which is incomplete in a different way
(Figure~\ref{fig:problem}). PIMS documentation provides broad coverage of
table schemas, column semantics, data types, and interface conventions, but
is organized by component rather than by planning task. Expert CASE records
provide sparse task-level evidence about which tables and fields experts
modify together, but omit schema definitions, business-language indexing,
and explicit explanations of why the modifications are coordinated. COM
specifications expose executable methods but do not encode refinery planning
logic. We use the term \emph{Documentation--Practice Gap} to describe this
source asymmetry observed in PIMS. The three sources are therefore
complementary rather than interchangeable: expert records can propose
task-level coordination hypotheses, documentation can ground them to valid
schemas and interfaces, and execution can reveal constraints that are not
explicitly stated in either static source. Consistent with this distinction,
raw documentation produces small and backbone-dependent gains, whereas
compiling the evidence into task-oriented skills improves performance
consistently.

We propose \textsc{SkillRefine}, a framework that addresses this gap through
role-separated cross-source skill induction and execution-driven refinement.
During skill construction, expert CASE records are used to propose
coordination patterns, while documentation grounds these patterns to valid
table schemas, column semantics, value constraints, and COM interfaces.
Interface stabilization separates failures in software invocation from
failures in business planning knowledge. The resulting skill library is then
refined through execution evidence: label-free format and solver-compliance
signals provide coarse failure screening, structured comparison with expert
references identifies table-, row-, column-, and value-level mismatches, and
tool-use trajectories are used as a localization heuristic to narrow the
candidate skill content to be repaired. The framework applies constrained
add, modify, remove, or constrain operations rather than rewriting the
library from a scalar score.

Importantly, the skill library is constructed and refined once using
DeepSeek-V4-Pro and is subsequently frozen and reused unchanged by all four
target LLM backbones. No backbone-specific skill editing or refinement is
performed on the held-out evaluation tasks. The evaluation therefore measures
both held-out task performance and whether the same skill artifact can be
reused across different agent backbones within the two evaluated PIMS
environments. It does not establish transfer to unseen refinery models,
production deployments, or other industrial planning platforms. Overall, we make the following contributions:

\begin{itemize}
  \item We introduce \textbf{role-separated cross-source skill induction for
  refinery planning software}. Sparse expert CASE records propose task-level
  coordination hypotheses, documentation grounds these hypotheses to valid
  table schemas, column semantics, value constraints, and COM interfaces, and
  execution evidence determines whether the resulting rules should be
  retained, constrained, or repaired.

  \item We formulate \textbf{execution-driven refinement as factorized
  diagnosis and localized repair over structured operations}.
  \textsc{SkillRefine} distinguishes format and solver-compliance failures
  from table-, row-, column-, and value-level mismatches, uses tool-use
  trajectories to narrow the candidate repair scope without claiming causal
  attribution, and applies add, modify, remove, or constrain operations to
  relevant skill content.

  \item We develop \textbf{PIMS-Bench}, an execution-grounded benchmark
  containing 140 planning tasks across five structural complexity levels and
  two PIMS demonstration models. Evaluation uses the actual COM and solver
  environment, expert-verified structured operation references, 100 held-out
  test tasks, and a single frozen skill library shared unchanged across four
  target LLM backbones.
\end{itemize}

\section{Related Work}
\label{sec:related_work}

\subsubsection{Agent Skill Acquisition and Refinement.}

Agent skills externalize reusable procedural knowledge so that agents can acquire task-specific behavior without parameter updates~\cite{agentSkillsSpec2026,xu2026agentskills}. Existing methods learn skills from different supervision: Voyager accumulates executable programs through exploration~\cite{wang2023voyager}; LGSD learns motor skills from language supervision~\cite{rho2025lgsd}; SkillRL uses skills for hierarchical reinforcement learning~\cite{skillrl2026}. Another line improves skills from execution feedback: Reflexion~\cite{shinn2023reflexion} and Self-Refine~\cite{madaan2023selfrefine} establish iterative self-improvement through verbal feedback; Trace2Skill converts trajectories into reusable skills with fix validation~\cite{ni2026trace2skill}; SkillOpt accepts edits through an execution-based validation gate~\cite{skillopt2026}; CREATOR~\cite{huang2024creator} disentangles skill creation from execution; SkillGrad optimizes via textual gradients~\cite{wang2026skillgrad}. Recent systems synthesize skills from external resources: SkillFoundry mines heterogeneous web evidence~\cite{shen2026skillfoundry}, SkillNet connects skills into a reusable network~\cite{liang2026skillnet}, and SkillForge combines knowledge bases with support tickets and failure diagnosis~\cite{liu2026skillforge}. SkillRefine differs by mining co-modification patterns from \emph{successful} expert records---rather than from failure cases---grounding inferred coordination rules against documentation schemas, and diagnosing failures at table, row, column, and value level.

\subsubsection{Implicit Procedures and Industrial Software Agents.}

Beyond skill acquisition, recovering undocumented procedures connects to tacit knowledge~\cite{polanyi1966}---knowledge that experts possess but cannot fully articulate. The core challenge in PIMS is inferring not only \emph{which} tables are co-modified but \emph{why}: expert CASE records reveal coordination patterns that documentation never describes, and SkillRefine extracts these through LLM-guided cross-source analysis (Phase~2a), grounding the inferred rationale against documentation schemas. LLM agents have been connected to optimization and engineering software: OptiGuide augments supply-chain solvers with natural-language what-if interaction~\cite{li2023optiguide}, OptiChat helps practitioners inspect and diagnose optimization models~\cite{optichat2025}, and a chemical engineering agent operates process simulators through model-context interfaces~\cite{liang2026llmagent}. These systems assume curated solver interfaces or pre-built model connections; SkillRefine instead learns the procedures that translate planning intent into coordinated table edits within an existing PIMS environment. The solver remains the authority for feasibility; its feedback refines the procedural layer.

\subsubsection{Skill Organization and Execution-Based Evaluation.}

Beyond individual skills, large skill libraries introduce two distinct risks: long-context degradation, where models underuse relevant evidence depending on its position~\cite{liu2023lost}, and skill shadowing~\cite{song2026shadowing}, where expanding the library increases selection errors even when individual skills are correct. Progressive disclosure~\cite{agentSkillsSpec2026}---loading skill metadata initially and full content on demand---mitigates loading cost. SkillRefine follows this separation and additionally records which skill was selected and which localized outcome followed, enabling joint revision of skill content and retrieval metadata. Evaluation benchmarks for agent skills include SWE-bench~\cite{jimenez2024swebench}, which evaluates code modification via GitHub issues, SkillsBench~\cite{li2026skillsbench}, which evaluates reusable skills across broad task suites, SWE-Skills-Bench~\cite{han2026sweskillsbench}, which studies repository-level skill acquisition and reuse, and SpreadsheetBench~2~\cite{zhu2026spreadsheetbench2}, which tests realistic workbook manipulation. PIMS-Bench differs in its unit of correctness: structured operations (table, row, column, value) over industrial planning models, with five complexity levels separating simple field edits from multi-object coordination procedures.

\section{Method: SkillRefine}

SkillRefine operates in four phases. \emph{Grounding} (Phase~1) constructs a structural and interface knowledge layer from documentation. \emph{Pattern extraction} (Phase~2) proposes candidate coordination hypotheses from expert CASE records and expands them with documentation. \emph{Interface stabilization} (Phase~3) ensures the COM execution layer is reliable. \emph{Structured refinement} (Phase~4) iteratively diagnoses and repairs skills through multi-signal execution feedback.

\begin{figure*}[t]
\centering
\includegraphics[width=\textwidth]{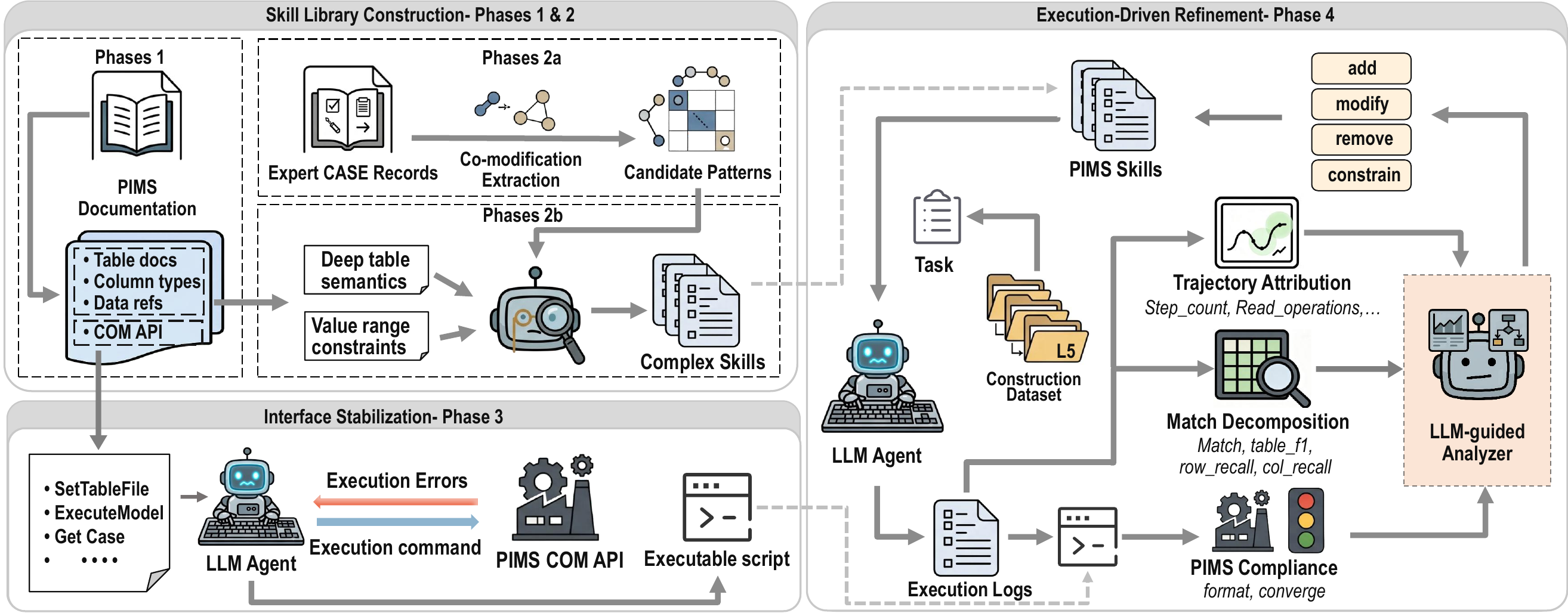}
\caption{SkillRefine pipeline. Phase~1 constructs a grounding layer from documentation. Phase~2 extracts candidate coordination patterns from expert CASE records and expands them with documentation. Phase~3 stabilizes the COM interface. Phase~4 refines skills through compliance screening, match decomposition, and trajectory-guided attribution.}
\label{fig:pipeline}
\end{figure*}

Skills are represented as standard \texttt{SKILL.md} documents following the Agent Skills specification~\citep{xu2026agentskills}: a front matter with name and description for progressive disclosure routing, a body with rules and procedures, and optional reference files for source evidence.

\subsection{Phase 1: Grounding Layer from Documentation}

Phase~1 constructs a grounding layer rather than task-level skills. PIMS documentation (3{,}126 HTML pages) is parsed into two structured outputs:
\begin{itemize}
  \item $\mathcal{S}_{\text{struct}}$: table schemas, column semantics (name, data type, valid range, required/optional flags), and inter-table data references for all 85 tables.
  \item $\mathcal{S}_{\text{interface}}$: COM automation method signatures (e.g., SetTableFile, ExecuteModel, GetObjectiveValue), parameter conventions, and execution state rules.
\end{itemize}
Parsing is rule-based: each table's HTML pages follow a consistent layout (description page, column reference page, optional example page), from which we extract table-level and column-level metadata into a unified JSON schema. Phase~1 does not generate business operation rules; it provides the schema and interface knowledge used to validate candidate patterns in Phase~2 and to ground skill content throughout the pipeline.

\subsection{Phase 2: Procedural Pattern Extraction and Schema-Grounded Expansion}
\label{sec:phase2}

Phase~2 builds a candidate skill library through two sub-phases.

\subsubsection{Phase~2a: Co-modification Extraction.}
We extract multi-table modification patterns from CASE records shipped with PIMS demonstration models: Weight Sample (6 non-empty records) and Gulf Coast Case Sets (317 records). For each CASE record $h_i$, we identify the set of tables modified together:
\begin{equation}
  C_i = \{T_1, T_2, \ldots, T_k\}
\end{equation}
Co-occurrence of $T_a$ and $T_b$ in $C_i$ does not imply that $T_a$ requires $T_b$; it indicates only that experts modified both in this scenario. The extracted patterns are therefore treated as \emph{candidate coordination hypotheses}, not as validated rules. The limited CASE coverage is not assumed to span the full task space; instead, CASE records seed a small set of task-level coordination hypotheses, while documentation supplies broad structural coverage and Phase~4 execution feedback determines which hypotheses are retained, revised, or removed. The construction tasks are manually selected to cover representative failure patterns rather than directly sampled from all CASE records.

\subsubsection{Phase~2b: Schema-Grounded Expansion.}
Phase~2a patterns cover at most 5 unique tables per model. Phase~2b expands coverage by reading documentation for tables not present in expert CASE records, performing three operations:
\begin{enumerate}
  \item \emph{Schema grounding}: verify that table names, column names, types, and value ranges in candidate patterns are valid against $\mathcal{S}_{\text{struct}}$.
  \item \emph{Intra-table rule extraction}: identify column interaction rules within a table (e.g., PROCLIM requires both MIN and MAX), value range constraints, and formatting rules.
  \item \emph{Cross-table hypothesis enrichment}: for tables that appear in Phase~2a co-modification patterns, associate documentation schema knowledge with the observed co-occurrence to produce candidate coordination explanations.
\end{enumerate}
All extracted rules are treated as hypotheses; their necessity is tested empirically through execution-based refinement in Phase~4.

\subsection{Phase 3: COM Interface Stabilization}

Phase~3 stabilizes the COM execution layer to prevent interface failures from being confounded with skill failures. All conditions use the same stabilized execution functions, including state reset, model backup/restore, solver invocation, and result extraction. Phase~3 introduces no business planning rules. All 50 executable test cases passed; 11 additional cases were skipped because the corresponding methods are unsupported in PIMS v22.0.11. 


\subsection{Phase 4: Structured Multi-Signal Refinement and Targeted Skill Repair}

Phases~1--3 produce a candidate skill library (47 documents, ${\sim}52$~KB) organized in a three-level hierarchy: L0 (meta-skills, ${\sim}200$ tokens: dependency awareness, format rules), L1 (domain skills, ${\sim}500$--1500 tokens: per-domain dependency graphs and operation rules), L2 (task skills, ${\sim}2000$ tokens: specific procedures for business scenarios). Progressive disclosure~\citep{xu2026agentskills} loads the skill index initially; full bodies and evidence files are loaded on-demand.

This initial library consistently improves performance over unguided baselines, but further improvement requires identifying and repairing specific skill deficiencies. We address this through \textbf{structured multi-signal refinement}: a framework that decomposes execution failures into factorized diagnostic dimensions and uses trajectory-guided attribution to identify candidate skills for repair.

\textbf{Three diagnostic layers.}

\emph{Layer 1: Compliance screening.} Two binary indicators screen for skill failures without requiring oracle answers:
\begin{equation}
  \mathbf{c}(d) = \bigl(f_d,\; v_d\bigr) \in \{0,1\}^2
\end{equation}
where $f_d$ is format validity (CASE syntax correctness: terminators, table keywords, column alignment) and $v_d$ is solver convergence (whether the LP solver accepted the modifications). A format failure indicates broken formatting guidance; a convergence failure indicates coordination rules that produce infeasible plans.

\emph{Layer 2: Structured match diagnosis.} Compliance does not imply correctness. When oracle answers are available, we factorize task correctness into operation-level dimensions:
\begin{equation}
  \mathbf{q}(d) = \bigl(m_d,\; t_d,\; r_d,\; c_d\bigr) \in [0,1]^4
\end{equation}
where $m_d$ is component match F1 (across table, row, column, and value tuples), $t_d$ is table F1, $r_d$ is row recall, and $c_d$ is column recall. Value correctness is included in component match F1 but is not separately reported as a diagnostic metric. Low $t_d$ suggests missing or over-generated table operations; low $c_d$ suggests incorrect operation-to-column mapping; low $r_d$ suggests incorrect entity-to-row mapping.

\emph{Layer 3: Signal-conditioned trajectory attribution.}
When Layer~1 or Layer~2 detects a failure, the recorded trajectory
\(
\tau_d =
\{
\text{consulted skills},\;
\text{tool sequence},\;
\text{reads},\;
\text{timing}
\}
\)
narrows the candidate repair scope. We derive two views from the
same trajectory:
\begin{equation}
    \tau_d^{\mathrm{comp}}
    =
    \Pi_{\mathrm{comp}}(\tau_d,c(d)),
    \qquad
    \tau_d^{\mathrm{match}}
    =
    \Pi_{\mathrm{match}}(\tau_d,e_d),
    \label{eq:conditioned_trajectory}
\end{equation}
where \(e_d=\operatorname{Diff}(\hat{a}_d,a_d^{*})\) is the structured
table-, row-, column-, and value-level mismatch. The first view
localizes candidate skills associated with format and solver failures,
whereas the second links structured errors to consulted skills. Both
are branches of a single attribution layer and are used only as
localization heuristics; they do not imply causality.

\textbf{Signal--diagnosis--repair mapping.} Table~\ref{tab:repair} maps observed signals to candidate diagnoses and repair actions.
\begin{table}[t]
\centering
\caption{Signal--diagnosis--repair mapping.}
\label{tab:repair}

\small
\setlength{\tabcolsep}{3pt}

\resizebox{\columnwidth}{!}{
\begin{tabular}{@{}lll@{}}
\toprule
\textbf{Signal} & \textbf{Diagnosis} & \textbf{Repair}\\
\midrule

Format invalid
& CASE/alignment error
& Modify format \\

Solver failure
& Missing dependency
& Add constraint \\

Low table F1
& Operation error
& Add/remove rule \\

Low row recall
& Entity--row mismatch
& Fix mapping \\

Low column recall
& Operation--column mismatch
& Fix mapping \\

Wrong skill
& Routing mismatch
& Update skill \\

Repeated reads
& Skill unclear
& Rewrite skill \\

Over-generation
& Rule too broad
& Restrict rule \\

\bottomrule
\end{tabular}
}
\end{table}

\textbf{Offline refinement process.}
Unlike prior methods that use bounded edits with a scalar validation
gate~\citep{skillopt2026} or trajectory-level consolidation without
factorized diagnosis~\citep{ni2026trace2skill}, our refinement uses
\emph{trajectory-guided attribution}. For each failing task, we
(1)~compute a structured diff when the output is parseable, or derive
a compliance diagnosis otherwise; (2)~associate the diagnosed error
with consulted skill documents through \texttt{read\_skill} tool-call
logs; and (3)~apply targeted repairs from Table~\ref{tab:repair}.
This process runs for $R{=}5$ rounds on
$\mathcal{D}_{\text{constr}}$:
\begin{equation}
\begin{aligned}
K_d &=
\operatorname{Trace}(\tau_d,c(d),e_d,\mathcal{S}),\\
\mathcal{S}' &=
\operatorname{Repair}(\mathcal{S},e_d,K_d),
\end{aligned}
\label{eq:repair}
\end{equation}
where $K_d$ denotes candidate skills localized through the two
signal-conditioned attribution branches. Each round makes
\emph{surgical corrections} rather than blind edits: for example,
a rule associated with column confusion (e.g., VPRICE instead of
PRICE) is corrected without rewriting unrelated skill content.
The output is a refined skill library $\mathcal{S}^*$ deployed to
the evaluation agent. The held-out test set
$\mathcal{D}_{\text{test}}$ is never used for refinement
(Algorithm~\ref{alg:skillrefine}).

\section{Experimental Setup}

\textbf{Models.} We evaluate on two AspenTech PIMS demonstration models: Weight Sample (10 crudes, 14 products, base objective 1036.12) and Gulf Coast (49 crudes, base 1341.02).

\textbf{Knowledge sources.}
We use 3,126 documentation pages from the PIMS CHM help
system and CASE records from the demonstration models
(6 non-empty Weight Sample records and 317 Gulf Coast
records) for procedural pattern extraction in Phase~2a.
Construction-task oracles supervise refinement in Phase~4.

\textbf{Benchmark.} PIMS-Bench comprises 140 tasks (70/model) across five complexity levels (28 each): L1 single-value changes, L2 multi-table coordination, L3 component-level modifications, L4 structural changes (including cross-model unit migration), and L5 system-wide coordinated modification. Tasks cover 31 tables and all four CASE keywords. They are partitioned into $\mathcal{D}_{\text{constr}}$ (40 tasks for skill construction and refinement) and $\mathcal{D}_{\text{test}}$ (100 held-out tasks). No test task participates in any skill construction or refinement decision. The tasks were jointly designed and reviewed by three refinery design experts.

\textbf{Conditions.} Four LLM backbones---DeepSeek-V4-Pro, DeepSeek-V4-Flash~\citep{xu2026deepseek}, Gemma-4-31B~\citep{gemma4_2026}, and Qwen-3.6-27B~\citep{yang2025qwen3}---are evaluated under six conditions using a ReAct-style agent~\citep{yao2022react}: \textsc{Baseline} (no external knowledge, 5 tools), \textsc{DocOnly} (+raw documentation, 6 tools), \textsc{Trace2Skill}~\citep{ni2026trace2skill} (trajectory-level patch induction and hierarchical consolidation with construction-set validation), \textsc{SkillOpt}~\citep{skillopt2026} (bounded edits with $D_{\text{constr}}$ validation gate), \textsc{SkillBase} (Phase~1--3 skills, unrefined, 7 tools), and \textsc{SkillRefine} (multi-signal refined skills, same 7 tools). \textsc{SkillBase} and \textsc{SkillRefine} share identical tools; the only difference is unrefined vs.\ refined skills. All conditions share the same PIMS execution interface; knowledge-enabled
conditions differ only in their documentation or skill access.

\textbf{Metrics.}
Primary metric is match F1 (component-level F1 across table,
row, column, and value tuples). For each task, the mean and
standard deviation are computed over five independent
executions and then averaged across the 100 held-out tasks.
Table F1 is used as an internal diagnostic signal during
refinement, while row and column recall are reported for
error analysis.

\textbf{Infrastructure.} All experiments run on a Windows~11 workstation (Intel Core i7-12700, 64~GB RAM; CPU only, no local GPU). PIMS v22.0.11 with the FICO Xpress v8.8.4 LP solver is operated via COM automation using Python~3.13 and \texttt{pywin32}. LLM backbones are accessed through an OpenAI-compatible API (temperature$=0.5$, max\_tokens$=8192$, native function calling with tool\_choice$=$auto). DeepSeek-V4-Pro and DeepSeek-V4-Flash support 1M-token context windows; Qwen-3.6-27B supports 32{,}768 tokens. Each backbone--condition pair comprises 500 executions (100 tasks $\times$ 5 runs, with a 50-step limit per run).

\begin{algorithm}[!t]
\caption{SkillRefine Pipeline}
\label{alg:skillrefine}
\small
\begin{algorithmic}[1]
\REQUIRE Documentation $\mathcal{D}$,
expert records $\mathcal{H}$,
construction tasks $\mathcal{D}_{\text{constr}}$,
construction oracles $\mathcal{O}_{\text{constr}}$,
PIMS execution environment $\mathcal{E}$,
refinement rounds $R$
\ENSURE Frozen refined skill library $\mathcal{S}^*$

\STATE $(\mathcal{S}_{\text{struct}},
\mathcal{S}_{\text{interface}})
\leftarrow \textsc{ParseDocs}(\mathcal{D})$

\STATE $\mathcal{S}_{\text{op}}
\leftarrow \textsc{ExtractPatterns}(\mathcal{H})$

\STATE $\mathcal{S}_{\text{op}}
\leftarrow \textsc{GroundAndExpand}
(\mathcal{S}_{\text{op}},
\mathcal{S}_{\text{struct}},
\mathcal{D})$

\STATE $\mathcal{S}_{\text{exec}}
\leftarrow \textsc{StabilizeInterface}
(\mathcal{S}_{\text{interface}},
\mathcal{E})$

\STATE $\mathcal{S}_0
\leftarrow \textsc{Encode}
(\mathcal{S}_{\text{struct}}
\cup \mathcal{S}_{\text{op}}
\cup \mathcal{S}_{\text{exec}})$

\STATE $\mathcal{S} \leftarrow \mathcal{S}_0$

\FOR{$r = 1$ to $R$}
  \STATE Execute $\mathcal{D}_{\text{constr}}$
  using $\mathcal{S}$; collect outputs, logs, and trajectories

  \FOR{each task $d_i$ in a fixed order}
    \STATE $c_i \leftarrow
    \textsc{Compliance}(\hat{a}_i,\ell_i)$

    \STATE $q_i \leftarrow
    \textsc{MatchDecomposition}(\hat{a}_i,a_i^*)$

    \IF{$c_i$ indicates failure or
    $\textsc{MatchF1}(q_i) < 1.0$}

      \IF{$\hat{a}_i$ is structurally parseable}
        \STATE $e_i \leftarrow
        \textsc{Diff}(\hat{a}_i,a_i^*)$
        over table, row, column, and value
      \ELSE
        \STATE $e_i \leftarrow
        \textsc{ComplianceDiagnosis}(c_i,\ell_i)$
      \ENDIF

      \STATE $K_i \leftarrow
      \textsc{Trace}
      (\tau_i,c_i,e_i,\mathcal{S})$
      \COMMENT{signal-conditioned localization; not causal}

      \STATE $\mathcal{S} \leftarrow
      \textsc{Repair}(\mathcal{S},e_i,K_i)$
      \COMMENT{add / modify / remove / constrain}

    \ENDIF
  \ENDFOR
\ENDFOR

\STATE $\mathcal{S}^*
\leftarrow \textsc{Freeze}(\mathcal{S})$

\RETURN $\mathcal{S}^*$

\end{algorithmic}
\end{algorithm}

\section{Results}

\begin{table*}[t]
\centering
\caption{Match F1 ($m$), Row Recall ($r_R$), and Column
Recall ($c_R$) across four LLM backbones and six conditions
on $\mathcal{D}_{\text{test}}$ (100 held-out tasks). For each
task, the mean and standard deviation are computed over five
independent executions; the table reports their averages across
the 100 tasks.}

\label{tab:distill}

\footnotesize
\setlength{\tabcolsep}{2.0pt}
\renewcommand{\arraystretch}{1.15}

\newcommand{\se}[1]{\scriptsize$\pm$#1}

\begin{tabular}{@{}l@{\hspace{4pt}}ccc@{\hspace{5pt}}ccc@{\hspace{5pt}}ccc@{\hspace{5pt}}ccc@{}}

\toprule

& \multicolumn{3}{c}{DS-V4-Pro}
& \multicolumn{3}{c}{DS-V4-Flash}
& \multicolumn{3}{c}{Gemma-4}
& \multicolumn{3}{c}{Qwen-3.6}
\\

\cmidrule(lr){2-4}
\cmidrule(lr){5-7}
\cmidrule(lr){8-10}
\cmidrule(lr){11-13}

& $m$ & $r_R$ & $c_R$
& $m$ & $r_R$ & $c_R$
& $m$ & $r_R$ & $c_R$
& $m$ & $r_R$ & $c_R$
\\

\midrule

Baseline & .581\se{.150} & .852\se{.036} & .818\se{.106} & .618\se{.144} & .832\se{.072} & .808\se{.091} & .620\se{.074} & .830\se{.031} & .684\se{.080} & .468\se{.174} & .829\se{.055} & .704\se{.149} \\

DocOnly & .620\se{.148} & .858\se{.034} & .795\se{.104} & .638\se{.142} & .824\se{.076} & .809\se{.102} & .653\se{.078} & .827\se{.034} & .716\se{.084} & .443\se{.182} & .817\se{.067} & .703\se{.155} \\

SkillBase & .694\se{.139} & .860\se{.050} & .871\se{.092} & .705\se{.137} & .835\se{.094} & .858\se{.086} & .694\se{.104} & .835\se{.050} & .821\se{.103} & .590\se{.185} & .835\se{.087} & .790\se{.130} \\

Trace2Skill & .747\se{.092} & \textbf{.884}\se{.027} & .889\se{.057} & .698\se{.123} & .832\se{.091} & .824\se{.128} & .771\se{.039} & .868\se{.025} & .875\se{.026} & .656\se{.117} & .842\se{.044} & .822\se{.090} \\

SkillOpt & .742\se{.078} & \textbf{.884}\se{.034} & .886\se{.060} & .719\se{.122} & .862\se{.073} & .861\se{.114} & .778\se{.040} & .872\se{.023} & .886\se{.048} & .711\se{.113} & .852\se{.043} & .861\se{.079} \\

\textbf{SkillRefine} & \textbf{.770}\se{.088} & .882\se{.053} & \textbf{.905}\se{.045} & \textbf{.757}\se{.079} & \textbf{.888}\se{.040} & \textbf{.897}\se{.042} & \textbf{.800}\se{.026} & \textbf{.893}\se{.012} & \textbf{.909}\se{.015} & \textbf{.764}\se{.072} & \textbf{.882}\se{.035} & \textbf{.897}\se{.035} \\

\bottomrule

\end{tabular}
\end{table*}

\subsection{Knowledge Extraction Results}

\textbf{Structural knowledge} (Phase~1): Documentation parsing extracted structured descriptions for all 85 PIMS tables, including column semantics, data types, and inter-table data references. Documentation provides 119 table-pair co-references (cascade-delete lists and data structure descriptions), but does not provide operational coordination guidance.

\textbf{Procedural knowledge} (Phase~2): Expert CASE records from PIMS demonstration models (Weight Sample: 6 non-empty records; Gulf Coast Case Sets: 317 records) show which tables experts modify together, but do not explain the underlying coordination rationale. Cross-source analysis with documentation schemas identifies candidate coordination rules grounded against column definitions and type constraints. Expert CASE records reveal co-modification patterns that encode how experts jointly modify tables, but cover at most 5 unique tables per model.

\textbf{COM interface stabilization} (Phase~3): The agent wraps 174 COM methods into Python callable functions and designs 61 test cases across 10 functional categories. Through 8 iterative rounds, all 50 executable test cases passed; 11 additional cases were skipped because the corresponding methods are unsupported in PIMS v22.0.11. \emph{None of the 140 PIMS-Bench tasks depend on skipped methods}. 

\textbf{Skill library statistics.}
Phases~1--3 produce an initial library of 47 skill documents (${\sim}52$~KB). After Phase~4 refinement, the library grows to 57 documents (${\sim}80$~KB): the analyzer identified failure patterns not covered by existing skills and synthesized 10 new rules. Under progressive disclosure, the agent initially loads only the skill index (${\sim}8.5$~KB); full bodies are loaded on-demand.

\subsection{Skill Conditions and Performance}
\label{sec:distillation-results}

Across the five difficulty levels, task complexity scales structurally: the average number of tables modified per task increases from 1.2 (L1) to 3.6 (L5), oracle modification components from 3.5 to 7.9, and hidden cross-table dependencies from 0.7 to 3.5, confirming that PIMS-Bench spans a meaningful range of planning challenges.

\begin{figure}[t]
\centering
\includegraphics[width=\linewidth]{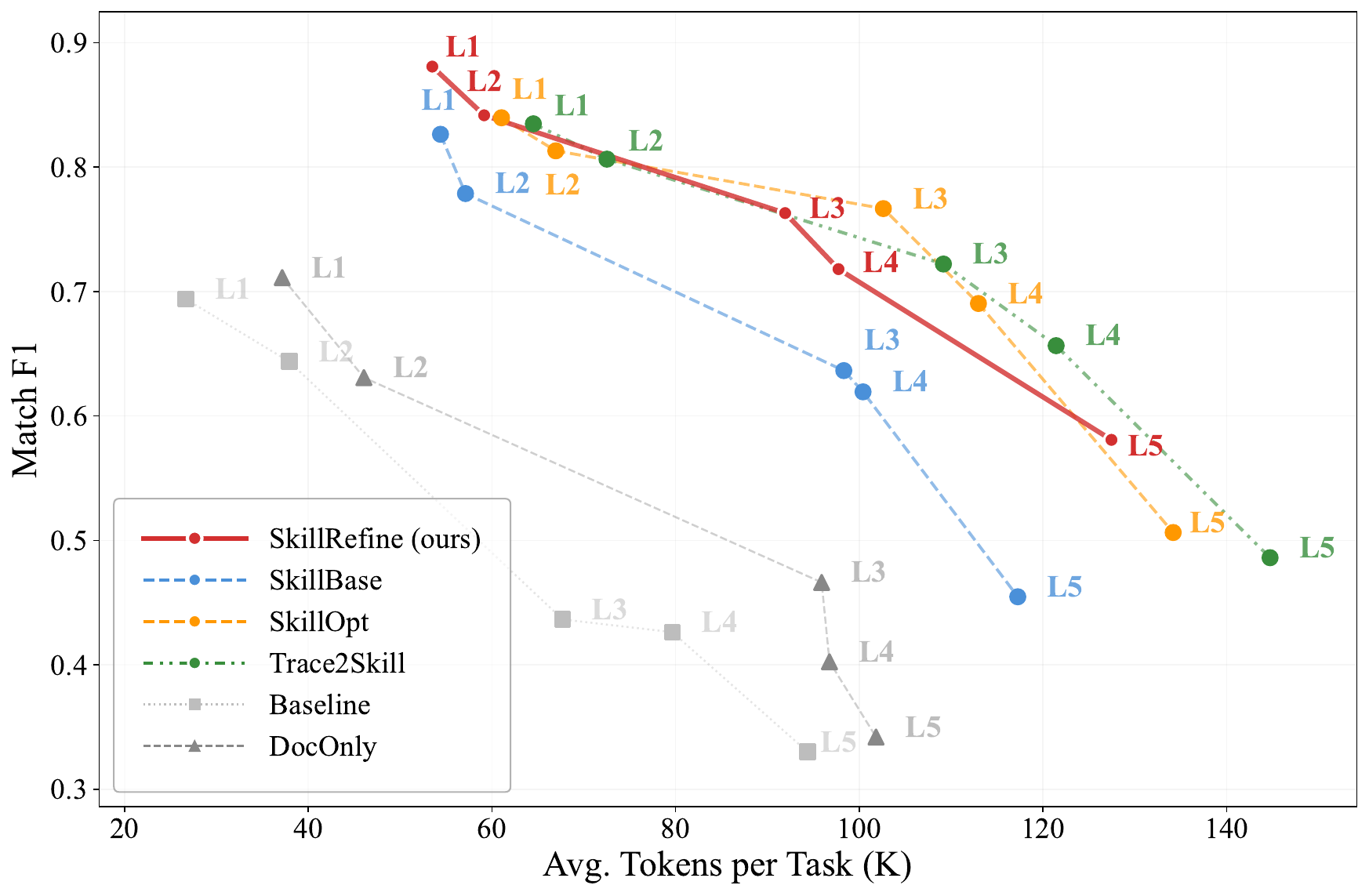}
\caption{Token efficiency versus accuracy across difficulty levels (L1--L5). Each line represents one condition; \textsc{SkillRefine} achieves the best accuracy--efficiency trade-off, with larger gains on harder tasks.}
\label{fig:efficiency}
\end{figure}

Table~\ref{tab:distill} reports results across four models and six conditions. Figure~\ref{fig:efficiency} visualizes the efficiency--accuracy tradeoff across difficulty levels.

\textbf{Raw documentation provides inconsistent gains.}
Relative to Baseline, DocOnly changes match F1 by +0.039,
+0.020, +0.033, and -0.025 across the four backbones. In
contrast, SkillBase consistently improves match F1 by
0.074--0.122, indicating that compiling evidence into
task-oriented skill packages is more effective than exposing
raw documentation. SkillRefine further improves SkillBase
by 0.052--0.174.

\textbf{External methods improve but fall short.}
\textsc{Trace2Skill} uses trajectory-level patch induction and
hierarchical consolidation, with construction-set validation for
skill selection, whereas \textsc{SkillOpt} applies bounded edits
through a scalar validation gate. Both improve over the baseline,
demonstrating the value of trajectory-derived knowledge and iterative
skill optimization. All refinement methods use the same construction
tasks and PIMS execution interface; they differ in how execution
feedback is converted into skill updates. However,
\textsc{Trace2Skill} lacks factorized error diagnosis, while
\textsc{SkillOpt}'s scalar signal does not identify which skill
content should be repaired. In contrast, \textsc{SkillRefine}
decomposes failures by compliance and structured match dimensions,
then uses signal-conditioned trajectories to localize targeted repairs.

\textbf{SkillRefine obtains the highest mean match F1 under
all four backbones.}
Compared with Baseline, SkillRefine improves match F1 by
0.189, 0.139, 0.180, and 0.296 on DeepSeek-V4-Pro,
DeepSeek-V4-Flash, Gemma-4-31B, and Qwen-3.6-27B,
respectively. Relative to SkillBase, the corresponding
improvements are 0.076, 0.052, 0.106, and 0.174. The
second-best method is Trace2Skill on DS-V4-Pro (0.747)
and SkillOpt on DS-V4-Flash (0.719), Gemma-4 (0.778),
and Qwen-3.6 (0.711).

\textbf{Column recall shows larger gains than row recall.} Among the reported diagnostic metrics, column recall shows larger gains than row recall on all four backbones, with improvements of $0.087$--$0.225$ over Baseline. This indicates that structured multi-signal refinement specifically helps agents identify correct column names---the most fine-grained operational knowledge. 

\textbf{The advantage grows with task complexity.} Averaged across the four backbones, the SkillRefine--Baseline gap increases from 10--15pp at L1--L2 to 18--28pp at L3--L5, confirming that cross-source skill induction provides the most value on complex multi-table coordination tasks.

\subsection{Ablation Study}
\label{sec:ablation}

\begin{figure}[t]
\centering
\includegraphics[width=\columnwidth]{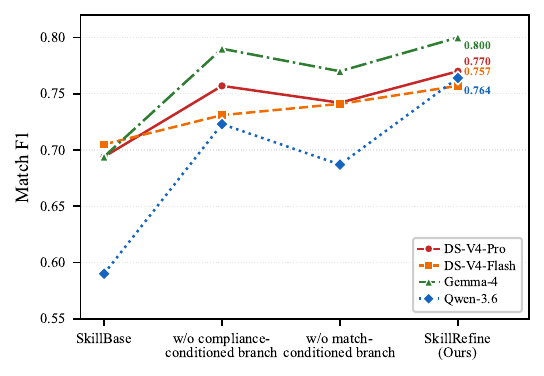}
\caption{Ablation of the two signal-conditioned
trajectory-attribution branches. Each variant removes the
corresponding localization branch while retaining its
diagnostic signal.}
\label{fig:ablation}
\end{figure}

We isolate the two signal-conditioned attribution branches.
In each ablation, the diagnostic signal remains available,
while only its trajectory-based localization is removed.
Figure~\ref{fig:ablation} reports match F1 across all four
backbones.

\textbf{Refinement is necessary.}
SkillBase, which uses cross-source extraction without
execution-driven refinement, achieves match F1 values of
0.590--0.705 across the four backbones. SkillRefine improves
these results to 0.757--0.800, corresponding to gains of
0.052--0.174, with the largest improvement on Qwen-3.6-27B.

\textbf{Both attribution branches contribute.}
Removing the compliance-conditioned branch reduces match F1
by 0.011--0.041, while removing the match-conditioned branch
reduces it by 0.016--0.077. The full model performs best on
every backbone, showing that the two branches provide
complementary localization evidence.

\textbf{Signal-conditioned trajectory attribution is a key
differentiator.}
Unlike methods that rely only on execution outcomes,
\textsc{SkillRefine} overlays tool-use context and consulted skills on
both diagnostic signals, helping identify repair candidates that the
outcomes alone cannot localize.

\section{Discussion}

\textbf{Cross-source induction is role-specific, not additive.} SkillRefine does not treat documentation and expert CASE records as symmetric retrieval corpora. CASE records provide sparse task-level evidence for proposing coordination patterns, while documentation supplies broad schema and interface grounding. Execution feedback then filters unsupported hypotheses and repairs incomplete rules. The contribution therefore lies in role-specific evidence composition rather than knowledge quantity alone.

\textbf{SkillBase improvement validates induction; SkillRefine validates refinement.} The improvement from Baseline to SkillBase (Table~\ref{tab:distill}) indicates that compiling heterogeneous evidence into task-oriented skills is already useful. The further improvement from SkillBase to SkillRefine shows that execution-driven correction is necessary to remove unsupported coordination rules, repair fine-grained column mappings, and constrain over-generalized procedures. These two gains separately support cross-source induction and execution-driven refinement.

\textbf{Model-dependent refinement sensitivity.}
SkillRefine improves over SkillBase by 0.076 on
DeepSeek-V4-Pro, 0.052 on DeepSeek-V4-Flash, 0.106 on
Gemma-4, and 0.174 on Qwen-3.6. The variation suggests
that different backbones differ in how effectively they use
localized skill corrections, although further evidence is
needed to explain this difference.

\textbf{Generalization scope.} This study evaluates held-out task generalization within two PIMS demonstration environments. The results support execution-driven skill refinement for structured industrial software within the evaluated environments. Generalization to unseen refinery models, production-scale deployments, or other industrial planning platforms remains to be established.

\textbf{Limitations.} (1)~Evaluation is limited to one industrial planning platform (AspenTech PIMS v22.0.11) and two demonstration models. (2)~Construction and test tasks share the same model environments and schema vocabulary, so the study evaluates held-out task generalization rather than transfer to unseen refinery models. (3)~Some planning tasks may admit multiple valid solutions, while the current structured match metric uses one expert reference. (4)~Trajectory-based error-to-skill attribution identifies plausible candidate skills but does not establish causal responsibility. 

\section{Conclusion}

We identified a Documentation--Practice Gap in PIMS, where component-oriented documentation provides broad structural coverage but lacks task-level coordination, while expert CASE records expose sparse operational patterns without explicit schema grounding. SkillRefine bridges this gap by using expert records to propose coordination patterns, documentation to ground them to valid schemas and interfaces, and execution evidence to diagnose and repair candidate skills. Its structured multi-signal refinement separates compliance
screening from table-, row-, column-, and value-level diagnosis and uses agent trajectories to narrow each repair to relevant skill documents. On 100 held-out tasks from PIMS-Bench, task-oriented SkillBase packages consistently outperform unguided baselines, and SkillRefine further improves the library, obtaining the highest mean match F1 under all four evaluated backbones. These results support execution-driven skill refinement for structured industrial software, while generalization beyond the evaluated PIMS environments remains to be established.

\bibliography{references}

\end{document}